\documentclass[aps,nofootinbib,superscriptaddress,twocolumn]{revtex4-1}

\usepackage{graphicx}
\usepackage{amsmath,amssymb}
\usepackage{hyperref}
\usepackage{braket}
\usepackage{booktabs}
\usepackage{subfigure}
\usepackage{subfloat}
\usepackage{float}
\usepackage[usenames]{color}

\hypersetup{
    colorlinks=true,
    linkcolor=red,
    citecolor=blue,
}

\allowdisplaybreaks

\def\be{\begin{equation}}
\def\ee{\end{equation}}
\def\ba{\begin{eqnarray}}
\def\ea{\end{eqnarray}}

\def\l{\left}
\def\r{\right}
\def\f{\frac}

\def\hub{{\mathcal H}}

\def\ie{{\frenchspacing\it i.e.}}

\begin{document}

\title{Priors on the effective Dark Energy equation of state in scalar-tensor theories}

\author{Marco Raveri} 
\affiliation{Kavli Institute for Cosmological Physics, Enrico Fermi Institute, The University of Chicago, Chicago, Illinois 60637, USA}
\affiliation{Institute Lorentz, Leiden University, PO Box 9506, Leiden 2300 RA, The Netherlands}
\author{Philip Bull} \affiliation{California Institute of Technology, Pasadena, CA 91125, USA}
\affiliation{Jet Propulsion Laboratory, California Institute of Technology, 4800 Oak Grove Drive, Pasadena, California, USA}
\author{Alessandra Silvestri} \affiliation{Institute Lorentz, Leiden University, PO Box 9506, Leiden 2300 RA, The Netherlands}
\author{Levon Pogosian} \affiliation{Department of Physics, Simon Fraser University, Burnaby, BC, V5A 1S6, Canada}

\begin{abstract}

Constraining the Dark Energy (DE) equation of state, $w_{\rm DE}$, is one of the primary science goals of ongoing and future cosmological surveys. In practice, with imperfect data and incomplete redshift coverage, this requires making assumptions about the evolution of $w_{\rm DE}$ with redshift $z$. These assumptions can be manifested in a choice of a specific parametric form, which can potentially bias the outcome, or else one can reconstruct $w_{\rm DE}(z)$ non-parametrically, by specifying a prior covariance matrix that correlates values of $w_{\rm DE}$ at different redshifts. In this work, we derive the theoretical prior covariance for the effective DE equation of state predicted by general scalar-tensor theories with second order equations of motion (Horndeski theories). This is achieved by generating a large ensemble of possible scalar-tensor theories using a Monte Carlo methodology, including the application of physical viability conditions. We also separately consider the special sub-case of the minimally coupled scalar field, or quintessence. The prior shows a preference for tracking behaviors in the most general case. Given the covariance matrix, theoretical priors on parameters of any specific parametrization of $w_{\rm DE}(z)$ can also be readily derived by projection.

\end{abstract}

\maketitle

\section{Introduction}

The cosmological constant $\Lambda$ is the simplest form of Dark Energy (DE) capable of driving the observed accelerating cosmic expansion \cite{Riess:1998cb,Perlmutter:1998np}. In General Relativity (GR), the bare (geometric) cosmological constant is a free parameter. However, $\Lambda$ also includes a constant vacuum energy density that receives contributions from the zero-point fluctuations of known particles \cite{Weinberg:1988cp}. To match the observed value, the geometric contribution to $\Lambda$ needs to be separately fine-tuned to compensate for the contribution of each different particle type individually, making the fine-tuning technically unnatural \cite{Burgess:2013ara}.

The extraordinary degree of tuning required to explain the observed value of $\Lambda$ within GR has spurred a broad search for mechanisms that can naturally set $\Lambda$ to a small value. A common proposal involves finding ways to prevent the vacuum from gravitating altogether, as in the so-called ``degravitation''  \cite{Dvali:2007kt}, ``self-tuning'' \cite{Charmousis:2011bf} or ``sequestering'' \cite{Kaloper:2013zca} scenarios, which would solve the naturalness problem at the cost of invoking some new mechanism -- generally a new field or force -- to cause the cosmic acceleration. These closely-related issues have motivated studies of alternative models of DE and modifications of gravity (MG). While not necessarily successful in resolving the fine-tuning problem themselves, many of these models can account for the acceleration, often predicting an effective DE that is dynamical. As the quest for a complete understanding of cosmic acceleration continues, it is of significant interest to know the extent to which DE can be dynamical and still remain consistent with observations.

Given the values of the Hubble parameter today, $H_0$, and the current matter density fraction, $\Omega_{\rm M}$, an arbitrary expansion history can be reproduced by assuming a flat Friedmann-Robertson-Walker (FRW) universe with a DE component that has an equation of state (EoS) $w_{\rm DE}$ with an appropriately chosen dependence on redshift \cite{Caldwell:1997ii}. Constraining $w_{\rm DE}(z)$, which we define here as the effective EoS of all non-dust contributions to the Friedmann equation at late times, is one of the primary science goals of ongoing and future surveys of large scale structure. A detection of $w_{\rm DE}(z) \ne -1$ would be strong evidence of new gravitational physics, and would constitute a potentially vital clue in understanding the source of cosmic acceleration.

Fitting a constant $w_{\rm DE}$ to current data results in good agreement with $-1$ \cite{Ade:2015xua}, but such fits would have missed subtle variations in $w_{\rm DE}(z)$, especially if the average happened to be close to $-1$. By the same token, using the Chevallier-Polarski-Linder (CPL) parameterization \cite{Chevallier:2000qy,Linder:2002et}, or any particular parametric form of $w_{\rm DE}(z)$, is prone to biasing the outcome \cite{Sahni:2006pa}. Forecasts using Principle Component Analysis (PCA) \cite{Crittenden:2005wj,Albrecht:2009ct} have shown that future surveys like Euclid and LSST will be able to constrain several eigenmodes of $w_{\rm DE}(z)$, justifying the use of more flexible parameterizations. 

A relatively general way to proceed is to use a piecewise representation, defining $w_{\rm DE}(z)$ in terms of its values in discrete bins in $z$. In the limit of many bins, fitting such a parametrization to data would avoid any bias, as any arbitrary $w_{\rm DE}(z)$ could be reproduced. In practice, however, if more than a few bins are used, many of them will be left unconstrained by the data, with values in neighboring bins effectively becoming degenerate. A possible way to remedy this problem is to introduce correlations between the bins by supplementing the likelihood with a prior covariance \cite{Crittenden:2011aa}. In this approach, the bias in the reconstructed $w_{\rm DE}(z)$ can be controlled for a given survey by using a PCA forecast. Specifically, as demonstrated in \cite{Crittenden:2011aa,Zhao:2012aw}, one can tune the strength of the prior to avoid the bias in the reconstructed evolution of $w_{\rm DE}(z)$ on timescales larger than a given correlation scale. A similar effect can be achieved using Gaussian Processes \cite{Shafieloo:2012ht,Seikel:2012uu}, which model correlations in the value of $w_{\rm DE}$ as a smooth function of redshift. Another approach, taken in \cite{Shafieloo:2012ht,Daly:2003iy,Shafieloo:2005nd}, attempts to derive $w_{\rm DE}(z)$ directly from data, by taking derivatives of the luminosity distances $d_{\rm L}(z)$. While seemingly model-independent, such methods require smoothing of the (noisy) data before derivatives can be taken, which effectively amounts to adopting a correlation prior. 

Since any attempt to constrain $w_{\rm DE}(z)$ will require a prior in one form or the other, one would like to develop reconstruction tools based on theoretically motivated priors that are explicit and easy to interpret.  Scalar fields (fundamental or effective) are ubiquitous in theoretical cosmology and particle physics, and are well-suited for representing a dynamical DE. A natural way to construct a theoretically-consistent, yet not-too-model-specific prior, is by generating a representative ensemble of scalar field models. Such approach was taken, for example, in \cite{Crittenden:2007yy,Marsh:2014xoa}, where priors on the CPL parameters $w_0$ and $w_a$ were derived based on an ensemble of very general quintessence DE models, \ie~minimally-coupled scalar fields \cite{Ratra:1987rm,Caldwell:1997ii}. In this paper, we go beyond minimally-coupled scalars, and consider the most general scalar-tensor models that are ``stable'', in the sense that they lack ghosts, gradient instabilities, and other pathologies. To generate the ensemble of models, we use the so-called ``unifying'' or ``effective field theory'' (EFT) approach to DE and MG \cite{Gubitosi:2012hu,Bloomfield:2012ff,Gleyzes:2013ooa, Bloomfield:2013efa,Piazza:2013coa,Gleyzes:2014rba,Gleyzes:2014dya}. Rather than working with $w_0$ and $w_a$, we derive the prior covariance for a binned $w_{\rm DE}(z)$, which makes it possible to project onto other parameterizations, and allows for reconstruction of $w_{\rm DE}(z)$ using the methods introduced in \cite{Crittenden:2011aa,Zhao:2012aw}. 

We generate the prior covariance of $w_{\rm DE}(z)$ for Horndeski theories, \ie~the general class of scalar-tensor theories with up to second order equations of motion \cite{Horndeski:1974wa,Deffayet:2011gz}. We also separately consider the widely studied sub-classes of the minimally couple scalar field, or quintessence \cite{Ratra:1987rm,Caldwell:1997ii}, and models with the canonical form of the scalar field kinetic energy, \ie~the generalized Brans-Dicke (GBD) models \cite{Brans:1961sx,carroll2004spacetime}. A non-minimally coupled scalar field mediates a fifth force which is tightly constrained in the Solar System. We will not concern ourselves with satisfying such constraints, as the models may include a screening mechanism that suppresses the fifth force in dense environments.  For our purposes, a scalar-tensor theory is viable if it has stable cosmological perturbations.

The paper is organized as follows. In Sec.~\ref{sec:eft} we introduce the EFT description of the cosmological background and linear perturbations in scalar-tensor theories and discuss a couple of representative forms of $w_{\rm DE}(z)$. We detail our method for sampling the space of viable Horndeski models in Sec.~\ref{sec:method}. In Sec.~\ref{sec:results} we present our results for the prior probability distribution for $w_{\rm DE}(z)$ and its covariance. We also provide an analytical form that accurately represents the correlation of $w_{\rm DE}(z)$ at any two points in $z$. We conclude with a summary in Sec.~\ref{sec:summary}.

\section{Modeling viable scalar field\\ Dark Energy}
\label{sec:eft}

We focus on the broad class of scalar-tensor models of gravity with second order equations of motion. 
The corresponding action in $(3+1)$ dimensions was derived by Horndeski~\cite{Horndeski:1974wa}, and later rediscovered in the context of generalized Galileons~\cite{Deffayet:2011gz}. For the purpose of our analysis, rather than working with the full covariant action of Horndeski gravity, it is convenient to employ the unifying framework, or effective theory approach (EFT),  formulated in~\cite{Gubitosi:2012hu,Bloomfield:2012ff} and further developed in~\cite{Gleyzes:2013ooa,Bloomfield:2013efa, Piazza:2013coa,Gleyzes:2014rba,Gleyzes:2014dya}. This allows us to model the background evolution and linear perturbations in a model independent way, in terms of a handful of free functions of time. The relevant EFT action reads
\ba
\mathcal{S} =&& \int d^4x \sqrt{-g}  \bigg\{ \frac{m_0^2}{2} \left[1+\Omega(\tau)\right] R + \Lambda(\tau) - c(\tau)\,a^2\delta g^{00}  \nonumber \\ 
+&& \frac{M_2^4 (\tau)}{2} \left( a^2\delta g^{00} \right)^2
 - \frac{\bar{M}_1^3 (\tau)}{2} \, a^2\delta g^{00}\,\delta {K}{^\mu_\mu}  \nonumber \\
+&&  \frac{\bar{M}_3^2 (\tau)}{2}\left[\left( \delta {K}{^\mu_\mu}\right)^2 - \delta {K}{^\mu_\nu}\,\delta {K}{^\nu_\mu}-\frac{ a^2}{2} \delta g^{00}\,\delta \mathcal{R}\right]+	\ldots \bigg\}\nonumber\\
+&& S_{m} [g_{\mu \nu}, \chi_m ],
\label{actioneft}
\ea
where $R$ is the four-dimensional Ricci scalar, $\delta g^{00}$, $\delta {K}{^\mu_\nu}$, $\delta {K}{^\mu_\mu}$ and  $\delta \mathcal{R}$ are respectively the perturbations of the upper time-time component of the metric, the extrinsic curvature, its trace, and the three dimensional spatial Ricci scalar of the constant-time hypersurfaces. The six functions, $\{\Omega,\Lambda,c, M_2^4,\bar{M}_1^3,\bar{M}_3^2\}$, are arbitrary functions of time allowed by the breaking of the time-diffeomorphism invariance, to which we refer to as the ``EFT functions''.  Finally,  $S_m$ is the action for all matter fields $\chi_m$ minimally coupled to the metric $g_{\mu \nu}$. Action (\ref{actioneft}) is built in the unitary gauge, where the additional scalar field is absorbed into the metric, and represents an expansion to quadratic order in perturbations around the flat Friedmann-Robertson-Walker (FRW) universe. A few extra terms were present in the original formulation of EFT of dark energy~\cite{Gubitosi:2012hu,Bloomfield:2012ff}, and different generalizations of~(\ref{actioneft}) have recently been studied in the literature~\cite{Kase:2014cwa,Frusciante:2015maa,Frusciante:2016xoj} to include models outside the Horndeski class. 
 
Perturbations in any specific Horndeski model can be mapped onto the action (\ref{actioneft}), with the correspondence between the EFT functions and the functions that appear in the full Lagrangian of the model given in \cite{Bloomfield:2013efa}. In the absence of a preferred scalar-tensor theory, one can adopt the agnostic point of view, and treat the EFT functions as free functions of time. We note that the GBD theories are represented by terms in the first line of~(\ref{actioneft}), while quintessence only requires specifying $\Lambda(\tau)$, with all other functions set to zero.

One of the advantages of using the EFT formulation is the ability to separate the terms that affect the background evolution from those that only concern the perturbations. In particular, only three of the EFT functions, $\Omega$, $c$ and $\Lambda$, play a role in determining the dynamics of the background, hence we will refer to them as the
\emph{background} EFT functions. Furthermore, as detailed in Appendix~\ref{App:background}, only two out of these three functions are sufficient to fix the background dynamics. In our approach, we specify $\{\Omega(a), \Lambda(a)\}$, and find the Hubble parameter, $\hub(a)$, by solving a differential equation:
\ba
\nonumber
\left(1+\Omega+\f{1}{2}a\Omega'\right)\f{d y}{d \ln a} &&+\left(1+\Omega+2a\Omega' +a^2\Omega'' \right)y \\
&&+\left( \frac{P_ma^2}{m_0^2} + \frac{\Lambda a^2}{m_0^2}\right) = 0\,,
\label{eq:hubble}
\ea
where $y \equiv \hub^2$,  and the prime indicates differentiation with respect to the scale factor. Given the solution for $\hub(a)$, the effective DE EoS is defined via
\begin{align}
w_{\rm DE} \equiv \frac{P_{DE}}{\rho_{DE}} = \frac{-2\dot{\hub} -\hub^2 - P_m a^2/m_0^2}{ 3\hub^2 - \rho_m a^2/m_0^2 } \ ,
\end{align}
where $\rho_m$ and $P_m$ are the combined energy density and the pressure of all particle species, and the over-dot denotes a conformal time derivative. The full details of this procedure are given in Appendix~\ref{App:background}. By sampling from a broad range of possible $\{\Omega(a),\Lambda(a)\}$ functions, as described in Sec.~\ref{sec:method}, we are able to generate a representative ensemble of possible expansion histories in single field DE models. We note that this procedure is different from the so-called designer approach used in e.g.~\cite{Hu:2014sea}, in which one provides $\hub$ (and $\Omega$), and solves for $\{c,\Lambda\}$. 

Since we are interested in reproducing the expansion histories of theoretically viable Horndeski models, we sample the functions $\{M_2^4,\bar{M}_1^3,\bar{M}_3^2\}$ along with $\{\Omega(a),\Lambda(a)\}$. While the former affect only the perturbations and, hence, would appear not to matter for the prior on $w_{\rm DE}(z)$, they do in fact play a role in determining whether a given background solution corresponds to a stable Horndeski theory. By perturbing the action (\ref{actioneft}) around a given background, one can derive constraints on combinations of EFT functions and their derivatives that exclude instabilities. Specifically, in our analysis, we impose the no-ghost and no-gradient-type instability conditions for scalar and tensor modes \cite{Frusciante:2016xoj}. We implement the procedure for solving for the background discussed above in EFTCAMB\footnote{\url{http://eftcamb.org}} \cite{Hu:2013twa,Raveri:2014cka} and use its stability module to filter out models with ghost and gradient type instabilities. We find that imposing these conditions leads to the model acceptance rate $\sim 50$\% for quintessence, $\sim 10$\% for GBD and   $\sim 1$\% for Horndeski, thus removing a noticeable fraction of the parameter space. The details on technical implementation of the stability conditions can be found in~\cite{Hu:2014oga}.

\subsection{Two simple case studies}
\label{sec:casestudy}

\begin{figure*}[htbp]
\centering
\includegraphics[width=\textwidth]{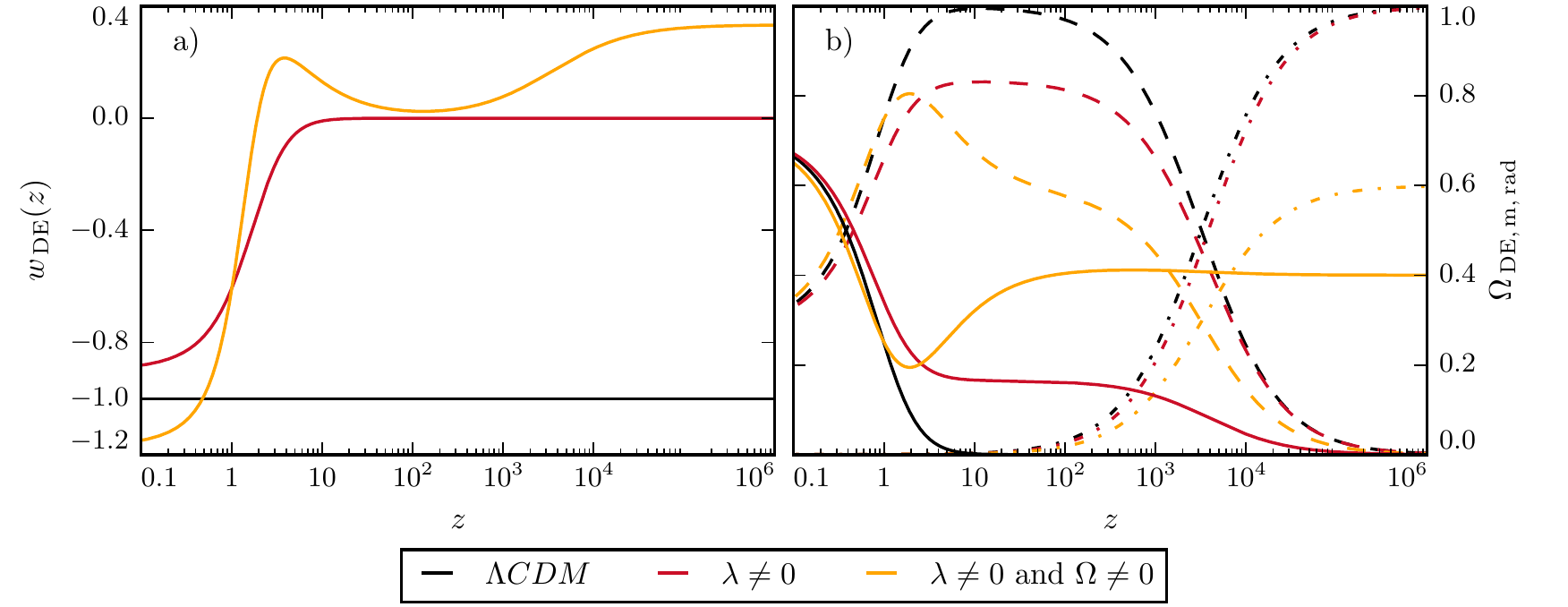}
\caption{The evolution of the effective DE EoS, $w_{\rm DE}(z)$ ({\it Panel a}) and the matter, radiation and effective DE densities ({\it Panel b}) in the $\Lambda$CDM model, in a model with a non-constant $\Lambda$, and a model with a non-constant $\Lambda$ and $\Omega \ne 0$.}
\label{fig:background} 
\end{figure*} 

To gain intuition into the expected behavior of $w_{\rm DE}(z)$ in scalar-tensor theories, let us examine the effect of varying $\Lambda$ and $\Omega$.

First, we consider a model in which all EFT functions except $\Lambda$ are set to their $\Lambda$CDM values. This choice corresponds to the minimally coupled quintessence models, with $\Lambda$ representing the Lagrangian of the scalar field, or the difference between its kinetic and potential energy densities. For our illustration, we take $\Lambda(a) = \Lambda_0\left( 1+ \lambda(a) \right)$, with $\lambda(a) = -0.1\,a$, and solve the background equations. Panel (a) of Fig.~\ref{fig:background} shows the DE EoS $w_{\rm DE}(z)$, while panel (b) shows the matter, radiation and DE energy fractions as a function of redshift. For this model, we see that $w_{\rm DE} \approx 0$ during the matter and radiation eras, before evolving towards $-1$ at low redshift. This behavior is quite generic, and can be understood as follows. The effective DE EoS is given by
\be
w_{\rm DE}(a)=\frac{\Lambda}{2c-\Lambda}= \frac{K-P}{\rho_{\rm DE}(a)}
\ee
where $K$ and $P$ are the quintessence kinetic and potential energy densities, and $\rho_{\rm DE} \equiv 3m_0^2H^2 - \rho_M - \rho_r$ is the total energy density excluding matter and radiation. In typical quintessence models, the field is slowly rolling at late times (low redshifts), with $P>K$, but retains a small non-zero kinetic term, hence  $w_{\rm DE} \gtrsim -1$. The energy density of the scalar field with $w_{\rm DE} \ge -1$ increases with redshift, unless the kinetic energy of the field is tuned to be exactly zero (which would be the case of the cosmological constant). This drives $w_{\rm DE} \to 0$ at high redshift, as seen in Fig.~\ref{fig:background}.

Next, let us consider a model in which both $\Omega$ and $\Lambda$ can vary in time. For illustration, we take $\Omega(a) = 0.2(a-1)/[1+0.7(a-1)]$ and $\lambda(a) = 0.3 + 1.0(a-1)$. As can be seen from Fig.~\ref{fig:background}, there are two key differences from the $\Lambda$-only model. Firstly, $w_{\rm DE}$ has a tracking behavior at early times, with $w_{\rm DE} \sim 0$ during the matter era and $w_{\rm DE} \sim 1/3$ in the radiation era. Secondly, $w_{\rm DE}$ crosses $-1$ at late times, when dark energy starts becoming important. Both features are quite representative of non-minimally coupled models. To understand the tracking behavior, observe that changing the value of $\Omega$ changes the gravitational coupling, while keeping the densities of the matter and the radiation species the same. In this case, the effective DE fluid compensates for the difference between the expected and the actual gravitational contributions of matter and radiation. The fluid must, therefore, take on an EoS that tracks the EoS of the dominant species at any given time.

The fact that the effective $w_{\rm DE}$ can cross $-1$ in non-minimally coupled models is well known \cite{Das:2005yj}. To see this, consider the form of the effective $w_{\rm DE}$ in the simple case of constant $\Lambda$ and $\Omega$,
\begin{align}
w_{\rm DE}(a) = \frac{\Lambda -\Omega P_{m}}{2c -\Lambda -\Omega \rho_m},
\end{align}
where $P_{m}$ and $\rho_m$ are the matter pressure and energy densities. It is clear from the new term in the denominator that the coupling to matter can act to decrease the energy density of the effective dark energy fluid, pushing $w_{\rm DE}$ below $-1$. While the expression for $w_{\rm DE}$ in the case of varying $\Lambda$ and $\Omega$ is more complicated, the physics is essentially the same.

\section{Methods}
\label{sec:method}

This section explains how we build the ensemble of viable Horndeski models, and the way in which we store and present the priors on $w_{\rm DE}(z)$.

\subsection{Sampling the ensemble of viable scalar field dark energy models}

A particular model realization is defined by specifying the EFT functions that appear in action (\ref{actioneft}). Ideally, one would adopt the most general functional forms possible for these functions, to avoid unduly focusing on only a few -- potentially unrepresentative -- corners of the space of possible models. In practise, this means choosing a parametrization that is broad enough to represent many different functions of time, while having sufficiently few parameters that it can be efficiently sampled using a Monte Carlo algorithm. To strike a balance between generality and tractability, we will use a number of different series expansions in the scale factor, $a$.

The first such expansion is a truncated Taylor polynomial, defined by
\begin{align} \label{eq:taylor_param}
f(a) = \sum_{n=0}^{N} \frac{\alpha_{n} }{n!} \left( a-a_0\right)^n \,,
\end{align}
where $N$ is the order at which we choose to truncate the expansion, $a_0$ is the point around which we expand, and $\{\alpha_n\}$ is a set of coefficients to be drawn from a random distribution. Given $\alpha_n$ coefficients with identical prior distributions, such a parametrization would favour the lower order terms.

The second expansion that we consider is a polynomial expansion,
\begin{align} \label{eq:poly_param}
f(a) = \sum_{n=0}^{N} \alpha_{n} \left( a-a_0\right)^n \,.
\end{align}
This differs from the Taylor expansion in the absence of the $n!$ term that suppresses higher order terms. Depending on how the $\alpha_n$ coefficients are drawn, this will allow rapid variations in $f(a)$ to arise more easily.

Finally, we will also use a Pad\'e expansion,
\begin{align} \label{eq:pade_param}
f(a) = \frac{\sum_{n=0}^{N} \alpha_{n} \left( a-a_0\right)^n}{1+\sum_{m=1}^{M} \beta_{m} \left( a-a_0\right)^m} \,,
\end{align}
where the truncation order is now set by both $N$ and $M$. This expansion is well suited for describing models in which the functions transition from one value at small $a$ to another at large $a$. 

We consider these three expansions around $a_0=0$ and $a_0=1$, to represent models that exhibit ``thawing'' and ``freezing'' behaviors respectively. We also explore several truncation orders, ranging from zero-order expansions, where all the three expansions reduce to constants, to ninth-order expansions that should allow very general model behaviors to be captured. We progressively raise the truncation order, obtaining results on the DE EoS prior for each order in turn. We find that the results stabilize beyond the fifth order, changing very little as the expansion order is further increased.

The prior distributions for all the coefficients $\{ \alpha_n\}$ and $\{ \beta_m\}$, that define particular realizations of the EFT functions, are chosen to be uniform in the range $[-1,1]$. We checked that extending the ranges of the flat priors does not affect the results. 

We have also adopted weak priors on cosmological parameters. The relative density of radiation is fixed by the CMB temperature, while the matter density today (the sum of the baryon and the CDM contributions) was drawn from the range $\Omega_{m}\in [0,1]$. We assume the universe to be flat and the sum of neutrino masses to be $\sum_{\nu} m_{\nu} = 0.06$ eV. The present-day DE density (a free parameter that is not fixed by flatness in the models we considered), was allowed to span $\Omega_{\rm DE}\in [0,1]$. The Hubble constant was allowed to vary in the range $H_0 \in [20,100]\,\mbox{km/s/Mpc}$. In addition, we impose a weak prior on the \emph{present} value of the gravitational coupling, allowing variations of no more than $10\%$, and the \emph{present} value of the speed of gravitational waves, allowing variations of no more than $20\%$. The latter two conditions help to exclude models that are in obvious contradiction with laboratory experiments and observations of the nearby universe \cite{Perenon:2015sla,Caves:1980jn, Moore:2001bv,Jimenez:2015bwa}.

When deriving the probability distributions of $w_{\rm DE}$ values at different $z$, we also take into account mild background data constraints to remove histories that are obviously ruled out. Namely, we use information from Baryon Acoustic Oscillation (BAO) measurements~\cite{Beutler:2011hx,Percival:2009xn, Padmanabhan:2012hf,Anderson:2012sa}, estimates of the Hubble constant~\cite{Riess:2011yx}, and supernovae distance measures~\cite{Suzuki:2011hu}, but with a significantly enlarged covariance (by a factor of four) to avoid the biasing of our results by tensions between these datasets. Let us stress that we do not use this information when deriving the covariance of the $w_{\rm DE}$ bins, so that it can be used in reconstructions of $w_{\rm DE}(z)$ as a purely theoretical prior.

We explore the space of cosmological parameters and $\{ \alpha_n\}$ coefficients using a Monte Carlo sampling procedure. For each sample, after solving the background equations (with initial conditions defined at the present day; see Appendix~\ref{App:background}), we check the stability of the corresponding model and, if accepted, compute $w_{\rm DE}(z)$. To ensure good coverage, we enforce a minimum number of $10^5$ accepted samples. Depending on the acceptance rate, this results in $\sim 10^6 - 10^8$ total samples.

While we have chosen several different ways to parametrize the behavior of the EFT functions, we can marginalize over these choices if desired. Given a set of models $\mathcal{M}$ corresponding to the different available options, we can compute
\begin{align} \label{eq:model_marginalization}
P(\vec{w}) = \sum_{\mathcal{M}} P\left( \vec{w}| \mathcal{M} \right) P\left( \mathcal{M} \right) \,,
\end{align}
where $P\left( \mathcal{M} \right)$ is the prior probability of each of the choices. Note that we are not interested in the normalization of $P\left( \vec{w}\right)$, which can be recovered a posteriori, so there is no need to impose the constraint $\sum_{\mathcal{M}} P(\mathcal{M})=1$. We will assume uninformative priors on the choice of parametrization, so that $P(\mathcal{M})$ is equal for all models. Since we have generated Monte Carlo samples of $P\left( \vec{w}| \mathcal{M} \right)$ with the same number of accepted points, we can then obtain $P(\vec{w})$ simply by merging the Monte Carlo samples from all parametrizations and re-weighting them with their respective acceptance rates.
 
\subsection{The priors on $w_{\rm DE}(z)$} \label{SubSec:wDEPrior}
\label{sec:wpriors}

Given an expansion history from a particular model realization, we store the corresponding $w_{\rm DE}(z)$ in $100$ linearly-spaced bins in scale factor that cover the redshift range $z\in [0,6]$. From the Monte Carlo samples of the binned EoS, $w_{\rm DE}(z_i)=w_i$, we compute the mean and the covariance,
\begin{align} \label{eq:sample_covariance}
\bar{w}_i &= \frac{1}{N_{\rm samp}}\sum_{\rm samples} w_i \,,\nonumber \\ 
C_{ij} &= \frac{1}{N_{\rm samp}-1}\sum_{\rm samples} \left( w_i - \bar{w}_i \right)\left( w_j -\bar{w}_j \right) \,.
\end{align}
One can also define the normalized correlation matrix as
\begin{align} \label{eq:sample_correlation}
\mathcal{C}_{ij} = \frac{C_{ij}}{\sqrt{C_{ii}C_{jj}}} \,.
\end{align}
Under the assumption of Gaussianity, the prior covariance matrix can be turned into a prior probability distribution and used in reconstructions of $w_{\rm DE}(z)$ from data, following the procedure suggested in \cite{Crittenden:2011aa}. 

While having a numerically-obtained discrete prior covariance matrix $C_{ij}$ may be sufficient for many practical applications, it is useful to also have an analytical expression characterizing the correlation of $w_{\rm DE}(z)$ between an arbitrary pair of redshifts. To this end, following \cite{Crittenden:2005wj,Crittenden:2011aa}, we can introduce a two-point correlation function $C(a, a')$ defined as
\begin{align} \label{eq:w_correlation}
C(a,a') \equiv \langle [w_{\rm DE}(a)-\bar{w}_{\rm DE}(a)][w_{\rm DE}(a)-\bar{w}_{\rm DE}(a')] \rangle \, .
\end{align}
As in the discrete case, Eq.~(\ref{eq:sample_correlation}), this prior covariance can be expressed in terms of the normalized correlation, $\mathcal{C}(a,a')$, equal to unity for $a=a'$, and the auto-correlation (or variance), $C(a) \equiv C(a,a)$, \ie
\begin{align} \label{Eq:ReconstructedCovariance}
C(a, a') = \sqrt{C(a)C(a')} \,\, \mathcal{C}(a,a') \,.
\end{align}
In Sec.~\ref{sec:results}, we will obtain analytical forms of $C(a, a')$ for different classes of models by separately fitting $\mathcal{C}(a_i,a_j)$ and $C(a_i)$ to the numerically-obtained $C_{ij}$ and $C_{ii}$. 

As derived quantities, in Sec.~\ref{sec:results}, we also compute projections of the binned $w_{\rm DE}(z)$ model onto parameters of the CPL parametrization,
\begin{align} \label{eq:CPLdefinition}
w_{\rm DE}(a) \approx w_0 + w_a(1-a) \,,
\end{align}
where $w_a$ is obtained from
\begin{align} \label{eq:CPLprojection}
w_a = - \left. \frac{dw_{\rm DE}}{da} \right|_{a=1}  \,.
\end{align}
This, amongst other things, allows us to compare our results to those in \cite{Marsh:2014xoa}, where priors on ($w_0$,$w_a$) were derived for quintessence by sampling a broad range of scalar field potentials.

\section{Results}
\label{sec:results}

\begin{figure*}[!htbp]
\centering
        \includegraphics[width=\textwidth]{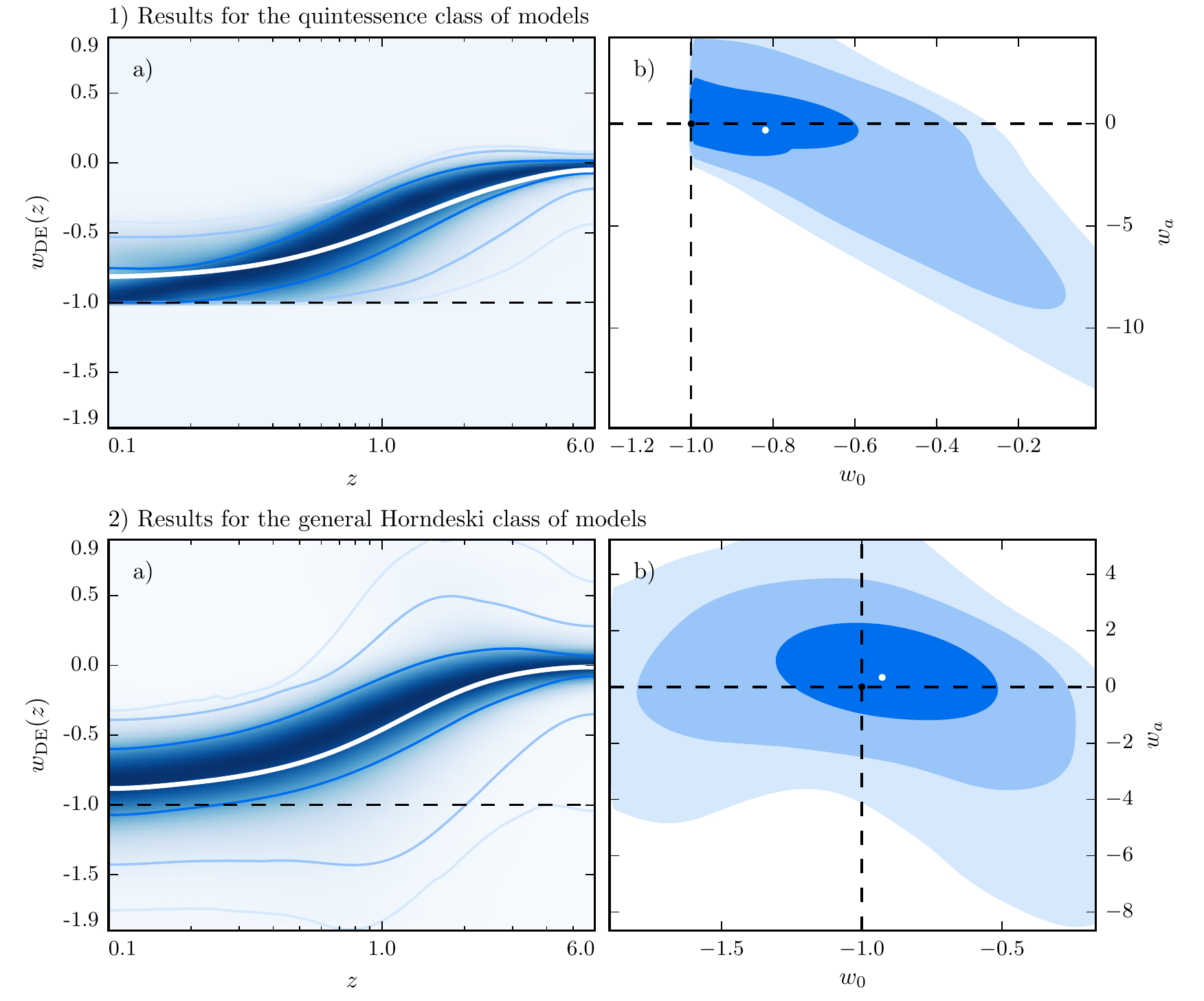}
    \caption{The ensemble of effective DE equations of state, $w_{\rm DE}(z)$, for the two model classes, with mild data constraints applied. {\it Panel a, all figures}: probability density of the effective DE EoS as a function of redshift. {\it Panel b, all figures}: probability density for the projection onto the $w_0$-$w_a$ parameters. In all panels, the white lines/points show the mean, and contours show the 68\%, 95\%, and 95\% C.L. intervals, the blue shading in the left-hand column shows the probability density, while the black point and dashed line represent the $\Lambda$CDM model.}
    \label{fig:eos}
\end{figure*}

In this section, we present the results of our Monte Carlo exploration of allowed DE EoS histories in scalar-tensor theories. We first discuss the PDFs of $w_{\rm DE}$ at different $z$, and then the covariance of $w_{\rm DE}(z)$ and the corresponding analytical fitting formulae.

\subsection{The ensemble of $w_{\rm DE}(z)$ functions}

By following the procedure for sampling the ensemble of scalar field theories and calculating the corresponding expansion histories described earlier, we obtain an ensemble of functions $w_{\rm DE}(z)$. 
Fig.~\ref{fig:eos} (left column) shows the probability density of $w_{\rm DE}$ as a function of $z$, derived from the Monte Carlo sample of two classes of models: the minimally coupled quintessence field, and the full Horndeski theory. These distributions have been marginalized over cosmological parameters, boundary conditions, EFT function parametrizations, orders of the Taylor expansion, and freezing/thawing behaviors (values of $a_0$). In addition, the mild observational constraints discussed in the previous section have been applied. The figures are shaded according to probability density, and show the mean $w_{\rm DE}(z)$, as well as the 68\%, 95\%, and 99\% C.L. contours.

Despite the large amount of freedom in the specification of the models, there is a remarkably well-defined structure to the $w_{\rm DE}(z)$ prior probability density. The overall behavior is similar between the two model classes, despite the significant increase in model freedom in going from the quintessence ($\Lambda$-only) class to the most general Horndeski class. In both cases, the EoS transitions from $w_{\rm DE} \approx 0$ deep in the matter-dominated regime to $\sim -1$ at low redshift. The high-redshift behavior is quite generic, for the reasons given in Sect.~\ref{sec:casestudy}: only quintessence models with kinetic energy fine-tuned to be close to zero can remain potential-dominated (and thus have $w \ll 0$) at early times, while the tracking behavior is common in non-minimally coupled models, giving $w_{\rm DE} \approx w_m = 0$ in the matter-dominated regime. 
The low-redshift behavior is typical of freezing models, where the kinetic energy of the scalar field always remains sub-dominant, and is significantly driven by the mild data constraints. These down-weight all of the $w_{\rm DE}(z)$ functions that result in non-accelerating cosmologies with $w_{\rm DE} > -1/3$ at $z=0$.

There are, however, some significant qualitative differences between the two model classes. As shown in panel~1a) of Fig.~\ref{fig:eos}, all models in the quintessence class respect $w_{\rm DE} \ge -1$ at all times, as expected. This bound is not respected in non-minimally coupled models, which have a significant probability density of models with $w_{\rm DE} < -1$ at low redshift. These ``phantom'' models are relatively common, with $40\%$ of viable Horndeski models having $w_{\rm DE} < -1$ at $z = 0$.

Another feature worthy of note is the lack of models with $w_{\rm DE} \simeq -1 = {\rm const.}$, \ie~a cosmological constant-like evolution at all redshifts. While values of $w_{\rm DE} \approx -1$ are very common at low redshift, they are rather unlikely in both model classes at $z \simeq 2$. This can be explained dynamically, by the tracking and kinetic energy tuning arguments above, but has interesting observational implications. While the fractional DE density is low at $z \gtrsim 2$, making measurements of the EoS more difficult, our results suggest that EoS constraints in this regime may actually be the most powerful for distinguishing a cosmological constant from dynamical DE.

The right-hand column of Fig.~\ref{fig:eos} shows the projection of the $w_{\rm DE}(z)$ curves onto the $w_0$ and $w_a$ parameters. The results for the quintessence class are shown in panel~1b). This should be compared with Fig.~1 of \cite{Marsh:2014xoa}, which also shows the ($w_0, w_a$) distribution for a Monte Carlo-generated sample of quintessence models. The orientation of the prior is roughly the same, with most models following a track that goes from $(w_0, w_a)=(-1, 0)$ towards more positive $w_0$ and more negative $w_a$. This is in part due to the $w_{\rm DE} \ge -1$ restriction that applies to quintessence models, which excludes a substantial region of the $(w_0, w_a)$ plane.

The distribution shown in the panel 1b) of Fig.~\ref{fig:eos} is quite broad, with a significant fraction of models having $w_a > 0$. In contrast, the models of \cite{Marsh:2014xoa} occupied a much narrower track that is almost entirely at $w_a < 0$. Part of the reason for this is that the Monte Carlo method in \cite{Marsh:2014xoa} used randomly-generated scalar field {\it potentials} to define the quintessence models, while we are working directly in terms of randomly-generated EFT functions. This effectively leads to a different weighting of the models; simple random distributions over the parametrized potentials may translate to non-trivial distributions over the EFT functions. In this case, the weighting obtained by parametrizing the theories at the EFT function level seems to make it easier to obtain thawing models (c.f. \cite{Huterer:2006mv}, who also find relatively more models with $w_a > 0$). Also, the models in \cite{Marsh:2014xoa} were selected to be ``physically-motivated,'' in the sense that the forms of their potentials were inspired by high-energy theory, like moduli fields and axions. These models form a more restricted class than the general, phenomenological EFT approach considered here. In particular, treating $\Lambda(a)$, which represents a particular evolution history for the quintessence Lagrangian, as being a priori arbitrary, amounts to a significantly more agnostic approach, leading to a broader allowed range of ($w_0, w_a$) values.

The $(w_0, w_a)$ distribution for the Horndeski class is broader and much less constrained, primarily as a result of the broadening of the $w_{\rm DE}(z)$ distributions, and the ability of the models to extend to $w_{\rm DE} < -1$. The bulk of the models remain within the vicinity of the $(-1, 0)$ point in both classes, although there is a very slight shift towards less negative $w_0$ and positive $w_a$ for the full Horndeski class. This is despite the fact that very few of the models have equations of state that resemble a cosmological constant out to high redshift. In fact, the mean functional form in the left-hand panels approximately satisfies $w|_{z=0} \to -1$ and $w|_{z\to\infty} \to 0$ in both model classes, which would imply $w_0 = -1$ and $w_a = +1$. These values are within the 68\% region for all three classes, but shifted slightly from the mean (shown by the white points in Fig.~\ref{fig:eos}). This is, at least in part, because both $w_0$ and $w_a$ have been evaluated directly at $z=0$. It is clear from Fig.~\ref{fig:eos}, however, that the derivative of $w_{\rm DE}(z)$ is not constant with redshift, \ie~the $(w_0, w_a)$ parametrization cannot fully reproduce its true redshift evolution when defined in this way.

In addition to considering the quintessence sub-class of Horndeski theories, we have separately studied the case of the GBD models, \ie~models with a canonical scalar field kinetic term. Within the EFT framework, they are described by $\Lambda(a) \ne {\rm const.}$ and $\Omega(a) \ne 0$, with $c(a)$ derived from $\{\Lambda, \Omega\}$, and the other functions set to zero. We find the results for the PDF of $w_{\rm DE}(z)$ in this case are practically the same as in the full Horndeski case. On one hand, this is expected, since only $\Lambda(a)$ and $\Omega(a)$ affect the background evolution. On other hand, the higher order functions figure in the stability conditions, which do affect the acceptance rate in the Monte Carlo sampling. We find, therefore, that the stabilization effect of higher order functions has a limited impact on the shape of the prior.

In the analysis presented so far, we have applied mild observational constraints to ensure that the models are broadly consistent with the properties of the real Universe (see Sect.~\ref{sec:method}). The results are qualitatively similar even if these constraints are relaxed, with the only significant difference being the presence of trajectories that remain above $w_{\rm DE}=-1/3$ at low redshift.

\subsection{Theoretical priors on correlations of $w_{\rm DE}(z)$}

\begin{figure*}[!htbp]
\centering
\includegraphics[width=\textwidth]{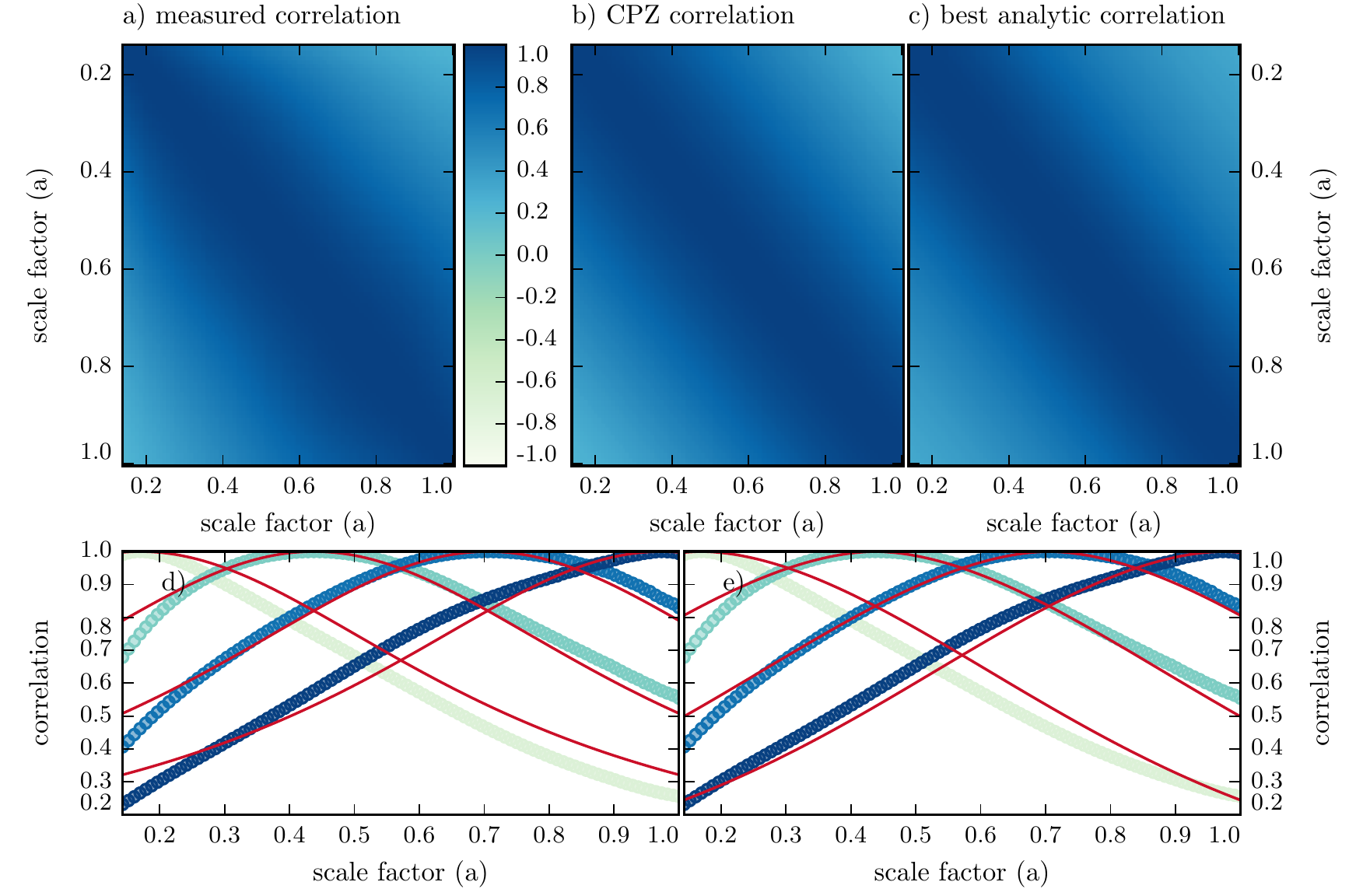}
\caption{Correlation results for the average over all quintessence models, without the mild data constraints.
{\it Panel a}: measured EoS correlation as a function of the scale factor. 
{\it Panel b}: best fit CPZ form  \cite{Crittenden:2005wj} (fixed power law) analytic prediction of the EoS correlation. See details in the text.
{\it Panel c}: best overall analytic prediction of the EoS correlation. See details in the text.
{\it Panel d,e}: correlation of the EoS at a single scale factor against all other scale factors, 
showing the measured correlation (dots) and the predicted correlation (continuous line) for the CPZ form and the best analytic model, respectively. Different colors correspond to different scale factors.
}
\label{fig:lambda_average_correlation}
\end{figure*}

\begin{figure*}[!htbp]
\centering
\includegraphics[width=\textwidth]{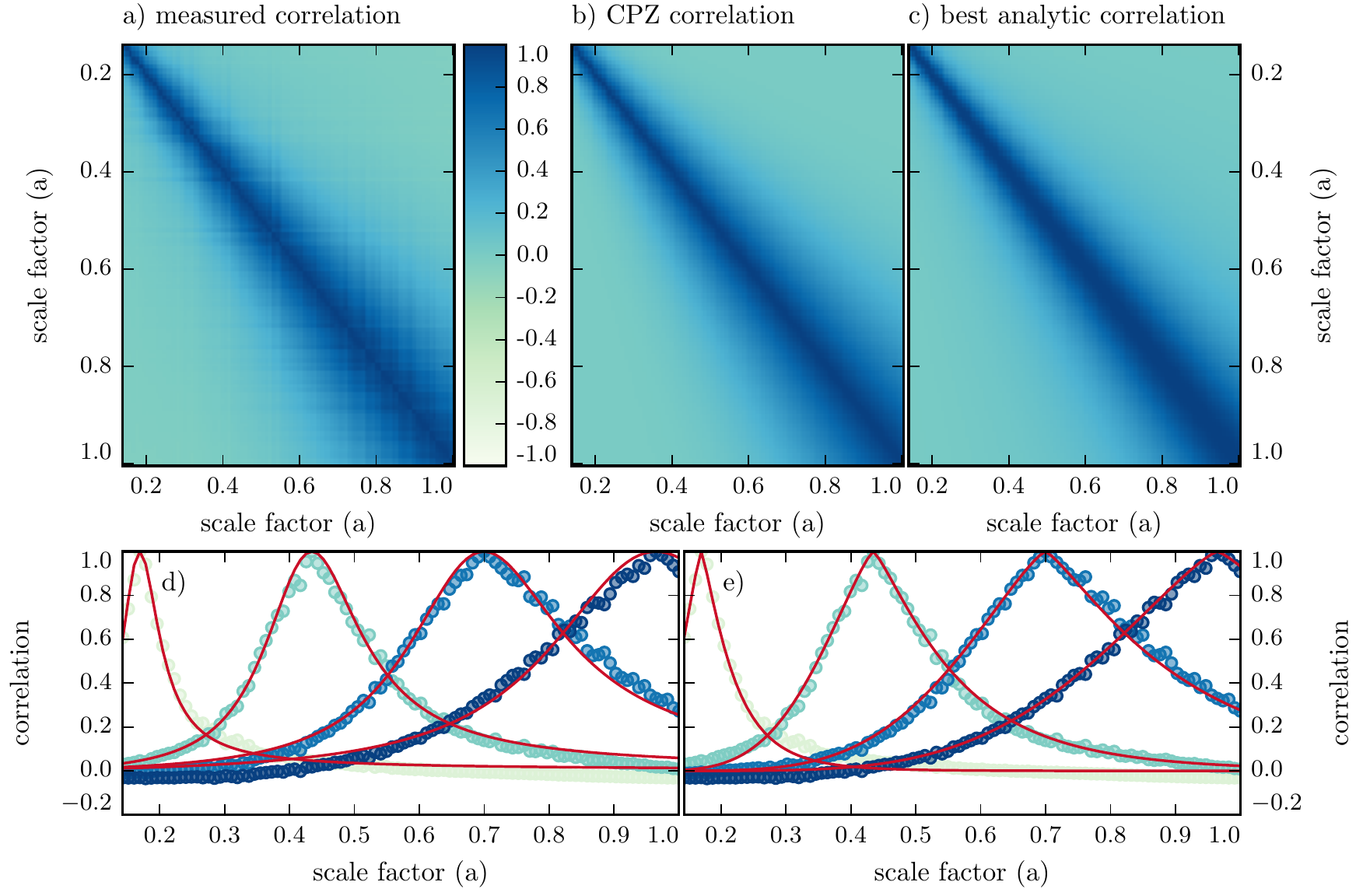}
\caption{Correlation results for the average over all Horndeski models, without the mild data constraints.
{\it Panel a}: measured EoS correlation as a function of the scale factor. 
{\it Panel b}: best fit CPZ form  \cite{Crittenden:2005wj} (fixed power law) analytic prediction of the EoS correlation. See details in the text.
{\it Panel c}: best overall analytic prediction of the EoS correlation. See details in the text.
{\it Panel d,e}: correlation of the EoS at a single scale factor against all other scale factors, 
showing the measured correlation (dots) and the predicted correlation (continuous line) for the CPZ form and the best analytic model, respectively.
Different colors correspond to different scale factors.
}
\label{fig:Hor_average_correlation}
\end{figure*}

\begin{figure*}[!htbp]
\centering
\includegraphics[width=\textwidth]{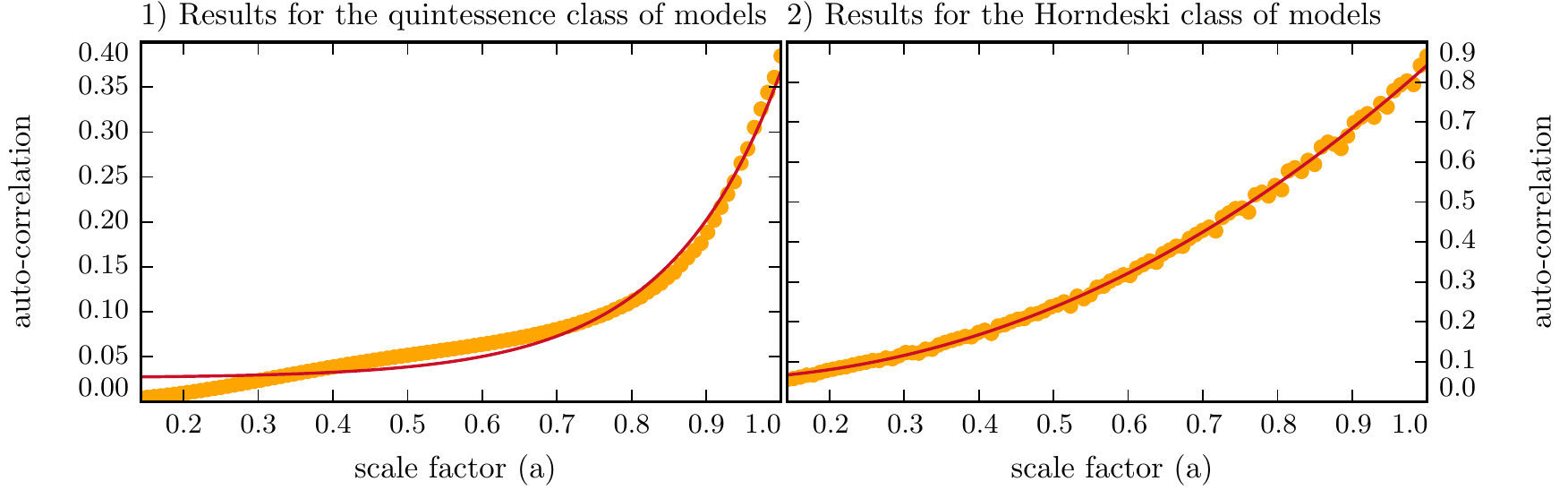}
\caption{Auto-correlation results for the quintessence (left panel) and Horndeski (right panel) model classes, without the mild data constraints. The dots represent the measured EoS auto-correlation while the continuous line is the best-fit parametrization.}
    \label{fig:autocorrelation}
\end{figure*}

In this section, we present the covariance between values of $w_{\rm DE}$ at different redshifts, obtained from averaging over the ensemble of $w_{\rm DE}(z)$ functions from each of the model classes. In doing so, we do not employ any observational constraints, as even relatively mild constraints result in correlations induced by the data overwhelming those coming from theory, effectively hiding the information that we want to extract.

We separate the covariance into the normalized correlation matrix, $\mathcal{C}(a_i,a_j)$, and the auto-correlation (or variance), $C(a_i)$, as in Eq.~(\ref{eq:w_correlation}). The top left panels in Figs.~\ref{fig:lambda_average_correlation} and \ref{fig:Hor_average_correlation} show the numerically obtained $\mathcal{C}(a_i,a_j)$ at binned values of $a$ for the quintessence and the general Horndeski classes, respectively. The points in Fig.~\ref{fig:autocorrelation} show the variance $C(a_i)$ for the two models.

As mentioned in Sec.~\ref{sec:wpriors}, having analytical expression for the correlation prior can simplify practical applications of our priors to reconstructing $w_{\rm DE}(z)$ from data using methods similar to those in \cite{Zhao:2012aw,Zhao:2017cud}. We can obtain them by fitting simple functional forms to the numerically obtained discrete covariance matrices. 

First, we perform a least-squares fit to the auto-correlation $C(a_i)$ using the following functional forms:
\begin{itemize}
 \item a Taylor expansion, $C(x) = \alpha +\beta\,(x-x_0)$;
 \item an exponential, $C(x) = \alpha  + \beta \, \exp[\gamma\,(x-x_0)]$;
 \item a power law, $C(x) = \alpha +\beta \, (x-x_0)^\gamma$;
\end{itemize}
where $x$ is the evolution variable for which we try three different choices: the scale factor, $x=a$; the redshift, $x=z$; and the number of e-folds, $x=\ln(a)$. We also try two choices of the reference point, $x_0$, in each case: for $x=a$ and $x=\ln(a)$, we either fix $a_0=1$ or allow $a_0$ to be free, and for $x=z$ we either fix $z_0 = 0$ or allow $z_0$ to be free. After fitting all of these different parametrizations to the measured auto-correlation function for each model class, we select the ones that result in the smallest residuals in each case. The choices that give the best fits are summarized in Table \ref{tab:data}, and plotted in Fig.~\ref{fig:autocorrelation} as a function of the scale factor. One can see that the exponential fitting formula with $x=\ln a$ and $a_0=1$ works very well in the Horndeski case, while the exponential form with $x=a$ and $a_0=1$ works best in the quintessence case. The analysis for the GBD class of models gives results that are almost indistinguishable from the Horndeski case. 

As expected, the auto-correlation (variance) is generally bigger in the Horndeski case, as these models permit more freedom than quintessence. In both classes, the auto-correlation is small at high redshifts, where the field is expected to have little or no dynamics, and so tends to result in similar values of $w_{\rm DE}$ in all models. The spread is much more significant at low redshifts however, where a greater range of dynamical behaviors is realized. The auto-correlation of the quintessence models is clipped by the hard bound of $w_{\rm DE} \ge -1$, reducing the allowed variation from model to model.

\begin{table}[!htbp]
\begin{center}
\begin{tabular}{l|cc}
\hline\hline
 & Best-fit auto-corr & residuals \\
\hline
Quintessence   & $0.03  +0.3 \, \exp[6.5 \,(a-1)]$ & $0.01$ \\
GBD               & $0.05  +0.8 \, \exp[1.8 \, \ln a ]$   & $0.007$ \\
Horndeski     & $0.05  +0.8 \, \exp[2 \, \ln a ]$      & $0.007$ \\
\hline\hline
 & Best-fit corr & residuals \\
\hline
Quintessence   & $\exp[-(|\delta a|/0.7)^{1.8}]$           & $9$ \\
GBD               & $\exp[-(|\delta \ln a|/0.3)^{1.3}]$      & $6$ \\
Horndeski     & $\exp[-(|\delta \ln a|/0.3)^{1.2}]$      & $6$ \\
\hline\hline
 & Best-fit corr (fixed CPZ) & residuals \\
\hline
Quintessence & $(1+(|\delta a|/0.6)^2)^{-1}$ & $11$ \\
GBD & $(1+(|\delta \ln a|/0.2)^2)^{-1}$ & $12$ \\
Horndeski & $(1+(|\delta \ln a|/0.2)^2)^{-1}$ & $13$ \\
\hline\hline
\end{tabular}
\end{center}
\caption{Summary of the auto-correlation and correlation function fits.}
\label{tab:data}
\end{table}

Next, we use a similar procedure to find the analytical form of $\mathcal{C}(a,a')$. We consider three fitting forms. One is a generalized version of the CPZ parametrization \cite{Crittenden:2005wj}:
\begin{align} \label{eq:PL_correlation_scale}
\mathcal{C}(x,y) = \frac{1}{1+\left(|x-y|/\xi \right)^n } \,,
\end{align}
where $x$ and $y$ are either the scale factor, the redshift or $\ln a$, $\xi$ is a parameter defining the correlation time scale, and $n$ is a free parameter. We also separately fit to (\ref{eq:PL_correlation_scale}) with $n=2$, which is the CPZ form. In addition, we consider an exponentially decaying correlation,
\begin{align} \label{eq:EXP_correlation_scale}
\mathcal{C}(x,y) = \exp \left[ -\left(|x-y|/\xi\right)^n  \right] \,,
\end{align}
where the physical meaning of the parameters is the same as in the previous parametrization (\ref{eq:PL_correlation_scale}). We then evaluate the least-squares distance between $\mathcal{C}(a_i,a_j)$ and the numerically found $\mathcal{C}_{ij}$, and find the parameters that minimize it, as well as the fitting form that results in the smallest residuals for each class of models.

We present the best fit analytical forms of $\mathcal{C}(a,a')$ and the corresponding residuals in Table \ref{tab:data}, along with best fit parameters of the CPZ parametrization. The two top right panels of Figs.~\ref{fig:lambda_average_correlation} and \ref{fig:Hor_average_correlation} show the best fit CPZ and overall best fit $\mathcal{C}(a,a')$ for the quintessence and Horndeski models, respectively. To have a more detailed visual check of the goodness of fit, in the bottom panel of Figs.~\ref{fig:lambda_average_correlation} and \ref{fig:Hor_average_correlation}, we also plot $\mathcal{C}(a,a')$ as a function of $a'$ for several fixed values of $a$. One can see that the exponentially decaying parametrization works marginally better than the CPZ form, as it has more freedom built in. Overall, the analytical expressions work well in reproducing the correlation in different models. We also note that the best fit correlation length, $\xi$, is roughly the same in both parametrizations for each model. 

One can clearly see that the correlation $\mathcal{C}(a,a')$ is long ranged for quintessence at all $a$, and is slowly decaying with $|a-a'|$. In Horndeski models, on the other hand, the correlation is generally shorter ranged, and decays faster at early times compared to late times. This is expected, since Horndeski models have more freedom in the choice of expansion histories compared to quintessence, hence values of $w_{\rm DE}$ at different times are less correlated.  

It is also interesting to note that the auto-correlation, $C(a)$, depends linearly on $a$ for quintessence, while it scales as $a^2$ in the Horndeski case. Also, the correlation $\mathcal{C}(a,a')$ scales as a power of $|a-a'|$ for quintessence, while in the case of Horndeski it scales as a power of $|\ln a - \ln a'|$. This is due to the fact that, in quintessence, $w_{\rm DE}$ is solely a property of DE, which is only minimally coupled to other fluids. In non-minimally coupled models, on the other hand, $w_{\rm DE}$ is an effective quantity, determined largely by how the other fluids scale with redshift. This is true especially at higher redshifts, where the variance and the correlation of $w_{\rm DE}(z)$ in Horndeski models is set by the dynamics of the matter fluid, which has a fixed time-dependence and uniformly distributed random amplitude set by $\Omega_M$.

\subsection{A note on practical applications of the prior}

As mentioned earlier, one can use the prior covariance obtained above to aid reconstruction of $w_{\rm DE}(z)$ from data using the method developed in \cite{Crittenden:2011aa} and applied in \cite{Zhao:2012aw,Zhao:2017cud}. There, a piece-wise (binned) $w_{\rm DE}(z_i) = w_i$ is fit to data along with other model parameters, $\vec{p}$, using the usual MCMC method for sampling the posterior PDF, ${\cal P}(\{w_i\},\vec{p})$. According to Bayes' theorem, the posterior PDF is the product of the likelihood and the prior PDFs:
\be
{\cal P}(\{w_i\},\vec{p}) \propto {\cal L}(\{w_i\},\vec{p}) \times {\cal P}_{\rm prior}(\vec{p}) \times  {\cal P}_{\rm prior}(\{w_i\})\ ,
\ee
where the prior on $\{w_i\}$ is obtained, under the assumption of Gaussianity, from the covariance matrix $C_{ij}$ discussed in the previous subsection:
\be
{\cal P}_{\rm prior}(\{w_i\}) \propto e^{-\sum_{ij} (w_i- {\bar w}_i) [C^{-1}]_{ij} (w_j- {\bar w}_j)/2} \ .
\label{eq:p_prior}
\ee
This prior probability explicitly depends on the average EoS, $\{\bar{w}_i\}$. We note that this is \emph{not} the mean $w_{\rm DE}(z)$ shown as white lines in panels 1(a) and 2(a) of Fig.~\ref{fig:eos}. The PDFs obtained in that figure used mild data constraints that pushed the mean closer to the observationally favoured $w_{\rm DE} \sim -1$ region.

In practice, \emph{any} reasonable choice for the fiducial ${\bar w}_i$, {\it e.g.} any choice within the 1$\sigma$ band of the EoS shown in Fig.~\ref{fig:eos}, should be more or less equally acceptable. The primary purpose of the prior is to add curvature to the PDF along otherwise flat directions in the parameter space, thus eliminating the degeneracies between $w_i$ and helping the MCMC to converge. A reconstruction that is highly sensitive to the choice of ${\bar w}_i$ should not be trusted.

Other options, discussed in \cite{Crittenden:2011aa}, include using the so-called ``running average'', or marginalizing over ${\bar w}_i$ altogether. We refer the reader to \cite{Crittenden:2011aa,Zhao:2012aw,Zhao:2017cud} for further details on the method and its application.

\section{Summary}
\label{sec:summary}

We have derived theoretical priors on  the effective DE EoS within the Horndeski class of scalar-tensor theories, which includes all models with a single scalar field that have second order equations of motion. We separately considered the widely studied sub-classes of the minimally coupled scalar field, or quintessence \cite{Ratra:1987rm,Caldwell:1997ii}, and models with the canonical form of the scalar field kinetic energy, \ie~the generalized Brans-Dicke (GBD) models \cite{Brans:1961sx,carroll2004spacetime}. Overall, we find that the covariance of $w_{\rm DE}(z)$ in GBD is indistinguishable from that of the general Horndeski case. 

Our priors on $w_{\rm DE}(z)$ are stored in the form of a covariance matrix for binned $w_{\rm DE}$, which can be projected onto priors on the parameters of any specific parametrization, such as CPL. 

We found that there are notable differences between the case of the minimally coupled quintessence and the non-minimally coupled models, both in the mean values and the covariance of binned $w_{\rm DE}$. We found simple analytical forms for the correlation function, describing the correlation of $w_{\rm DE}$ at different redshifts, that fit our numerical results well. These should simplify the practical application of the priors to reconstructions of $w_{\rm DE}(z)$ from data. Such a reconstruction will be the subject of future work.

\acknowledgments

We thank Wayne Hu and Gong-Bo Zhao for useful comments.
MR is supported by U.S. Dept. of Energy contract DE-FG02-13ER41958. PB is supported by an appointment to the NASA Postdoctoral Program at the Jet Propulsion Laboratory, California Institute of Technology, administered by Universities Space Research Association under contract with NASA. AS acknowledges support from the NWO and the Dutch Ministry of Education, Culture and Science (OCW), and also from the D-ITP consortium, a program of the NWO that is funded by the OCW. The work of LP is supported by the Natural Sciences and Engineering Research Council of Canada (NSERC).

\appendix

\section{The background solution in EFT} \label{App:background}
Here, we discuss in detail the procedure for solving for the cosmological background evolution given a choice of EFT functions. In what follows, the over-dot represents a derivative with respect to the conformal time, the accent mark represents a derivative with respect to the scale factor, and the subscript $m$ indicates the sum over all particle species: CDM, baryons, photons, and massless and massive neutrinos. The Friedmann equations for EFT are given by:
\ba \label{Eq:Fr1}
\hub^2=&&\f{a^2}{3m_0^2(1+\Omega)}(\rho_m+2c-\Lambda)-\hub\f{\dot{\Omega}}{1+\Omega},\\ \nonumber
\dot{\hub}=&&-\f{a^2}{6m_0^2(1+\Omega)}\l(\rho_m+3P_m\r) \\
\label{Eq:acc1}
&&-\f{a^2(c+\Lambda)}{3m_0^2(1+\Omega)}
-\frac{\ddot{\Omega}}{2(1+\Omega)} \,.
\ea
Changing the time coordinate from conformal time to the scale factor, Eq.~(\ref{Eq:Fr1}) can be recast as
\begin{align} 
\label{eq:cConstraint}
& \frac{ca^2}{m_0^2} = \frac{3}{2}\left( 1+\Omega+a\Omega' \right)\hub^2 -\frac{1}{2}\frac{a^2\rho_m}{m_0^2} +\frac{1}{2}	\frac{\Lambda a^2}{m_0^2} \,.
\end{align}
Combining this with Eq.~(\ref{Eq:acc1}), and introducing $y \equiv \hub^2$, we obtain
\ba
\nonumber
\left(1+\Omega+\f{1}{2}a\Omega'\right)\f{d y}{d \ln a} &&+\left(1+\Omega+2a\Omega' +a^2\Omega'' \right)y \\
&&+\left( \frac{P_ma^2}{m_0^2} + \frac{\Lambda a^2}{m_0^2}\right) = 0 \ ,
\label{eq:hubble_app}
\ea
which is the differential equation we solve to find $\hub(a)$. The initial conditions for this equation, along with the current values of the model parameters, can be derived from the Friedmann constraint (\ref{Eq:Fr1}) computed today,
\ba \label{Eq:FlatnessConstraint}
&& \f{c_0}{m_0^2} = \f{3}{2}\hub_0^2\left( 1+ \Omega_0 +\Omega'_0 -\Omega_m^0 -\Omega_{\Lambda}^0\right) \,,
\ea
where we have defined
\ba
&& \Omega_m^0 \equiv \f{1}{3}\f{\rho_m^0}{\hub_0^2m_0^2} \hspace{0.5cm};\hspace{0.5cm} \Omega_{\Lambda}^0 \equiv -\f{1}{3}\f{\Lambda_0}{\hub_0^2m_0^2} \,.
\ea
A few comments regarding this equation are in order. The definition of matter densities follows from what we can directly measure, at least in principle. The quantities $\rho_{m}^0$ are in fact the densities that can be measured with a non-gravitational experiment by an observer that is assuming that all these species are moving on geodesics of the metric.
These quantities are combined in the equation (\ref{Eq:FlatnessConstraint}) that is usually used as the flatness constraint. Normally, when assuming flatness, we use this to compute the value of $\Omega_{\Lambda}^0$ given the value of $\Omega_{m}$. Here, the situation is slightly different, as both $\Omega_{\Lambda}^0$ and $\Omega_{m}^0$ can be chosen arbitrarily, and constitute two of the theory parameters that we are going to sample.
On the other hand, the flatness constraint is satisfied, once the present day value of the gravitational constant is fixed, by a suitable choice of the present day value of $c$.

Once we have a solution of Eq.~(\ref{eq:hubble}), we can deduce the effective DE pressure and density from the standard form of the Friedmann equations:
\begin{align}
\hub^2 &= \frac{a^2}{3m_0^2} \left( \rho_m +\rho_\nu +\rho_{DE} \right) \,, \nonumber \\
\dot{\hub} &= -\frac{a^2}{6m_0^2}\left( \rho_m +\rho_\nu +\rho_{DE} +3P_m +3P_{\nu} +3P_{DE} \right) \,,
\end{align}
and compute the effective DE EoS as:
\begin{align}
w\equiv \frac{P_{DE}}{\rho_{DE}} = \frac{-2\dot{\hub} -\hub^2 - P_m a^2/m_0^2}{ 3\hub^2 - \rho_m a^2/m_0^2 } \ .
\end{align}

\bibliography{eft-w}

\end{document}